\documentclass{aastex}
\usepackage{spr-astr-addons}
\usepackage{url}

\RequirePackage{color}

\newcommand{\emaila}{gpace@astro.up.pt}

\begin{document}

\title{The  discontinuous nature of  chromospheric activity evolution}
\slugcomment{HELAS--NA5 workshop,  Rome, 22--26 June 2009}
\shorttitle{The  discontinuous nature of  chromospheric activity evolution}
\shortauthors{Pace}

\author{G. Pace\altaffilmark{1}}
\affil{Centro de Astrofisica da Universidade do Porto}
\email{\emaila}

\begin{abstract}  Chromospheric  activity has  been  thought to  decay
  smoothly  with  time and,  hence,  to  be  a viable  age  indicator.
  Measurements in solar type stars in open clusters seem to point to a
  different  conclusion:  chromospheric   activity  undergoes  a  fast
  transition from Hyades level to that of the Sun after about 1 Gyr of
  main--sequence lifetime and any  decaying trend before or after this
  transition  must  be  much  less  significant than  the  short  term
  variations.
\end{abstract}


\section*{Introduction}
\label{sec:intro}

In the atmosphere  of the Sun and of the stars  similar to it, several
processes  take place, which  make the  temperature of  the atmosphere
higher  than  it  would   be  if  radiative  equilibrium  held.   Such
non--radiative  heating  mechanisms  are  powered  by  convection  and
magnetic field. In the outer layers of these stars, the temperature is
increasing  towards the  surface, and  the main  cooling  mechanism is
radiative loss through  strong resonance lines, such as  Ca {\sc II} H
and K,  Mg {\sc II}  h and k,  H$_{\alpha}$.  \citep[see e.g.][  for a
review]{hallrev}.  This region  is  referred to  as chromosphere,  the
radiation emitted  in it  is called chromospheric  activity (hereafter
CA),  and  the  strength  of   the  emission  cores  in  some  of  the
aforementioned resonance lines are a good CA indicator.

Solar--type  stars loose  angular momentum  due to  magnetised stellar
winds, and CA  is believed to decay with time  as a consequence.  This
trend is superimposed to short  term CA variations, such as those that
caused the Maunder--minimum in  the Sun, activity cycles equivalent to
the 11--year long solar one, and, in the most active stars, rotational
modulation \citep{wilson,skuma,noyes,simon}.

Due to its decay, CA is  a potential age indicator and several efforts
have been undertaken  in order to calibrate it  \citep[see e.g.]  [and
references  therein]{sod91,lach99,MH}.  But  results  coming from  the
analysis of solar type stars in open clusters are challenging the view
of a smooth, decreasing trend modulating CA cycles. We do not question
the main picture described above on CA decay, but it appears that this
occurs   mainly  in   one  major   event,  in   which  it   spans  the
Vaughan--Preston gap, and any evolution before and after such event is
less important than the variation due to activity cycles.

Rotation is  probably better  correlated with stellar  ages \citep[see
e.g.][]{gyrochron,cc07},  since  it is  not  affected  by any  cyclic
variation.   However, rotation periods  measurements are  not possible
for inactive  stars, for which we  can only have an  estimation of the
projected rotation  velocity, which is  not sufficient to to  draw any
firm conclusion on the nature of the angular momentum evolution.

The matter  deserves close attention, because  chromospheric ages have
been  used  in  important  studies  on chemical  enrichment  and  star
formation  in the Galactic  disk \citep{rp00a,rp00b},  and the  age of
stars  hosting planetary  systems  \citep{saffe}.  We  should reach  a
definitive conclusion  on whether CA is  or not a  good age indicator,
and this can be better achieved by observing more solar--type stars in
open clusters and by determining more precisely their ages. 

Intense investigation has long been conducted on the topic, especially
by using the  large amount of data collected  in Mount Wilson campaign
\citep{MWC}  and in the  planet-search surveys  \citep{pss}.  However,
only the  possibility of  observing at high  signal to noise  and high
resolution solar type stars in old and distant open clusters, provided
by the 8-- and 10--m class  telescopes, could make us progress in this
field, and it is the key to address still unanswered questions.

In  the present work,  I will  report on  the results  that led  us to
reconsider the viability of CA as age indicator. It is also shown that
data  used  to calibrate  chromospheric  ages  do  not contradict  our
conclusions.

\section{Data sample}
\label{datasample}
The complete sample on which  our conclusions are based consists of 40
main--sequence stars  in 7 open clusters  and the Sun.   The nature of
single, dwarf members of the parent clusters were established, for the
40  target   stars,  in  published   photometric,  proper--motion  and
radial--velocity studies \citep{hs,lmmd92,naa96,man02,mm90,tatm}, and,
only  for   IC~4756  and  NGC~5822,  from   our  own  radial--velocity
measurements \citep{papoc2}.   The data about  15 stars in  the Hyades
cluster  consist of  spectra taken  with  HIRES at  Keck, kindly  made
available to us  by D.  Paulson, A.  Hatzes and  P. Cochran.  The rest
of  the  data consists  of  spectra  taken with  UVES  at  VLT in  two
different runs.  In  the first (ESO run 66D-0457,  P.I.  L.  Pasquini)
the targets were: 7 in Praesepe, 2 in NGC~3680, 5 in IC~4651, and 6 in
M~67.  In  the second run  (ESO run 73D-0655,  P.I.  G. Pace)  2 stars
were observed in NGC~5822 and 3 in IC~4756.

Our  UVES  spectra have  a  resolution  of  R$\approx$100\,000 in  the
spectral range  from 4\,800 to 6\,800 {\AA},  and R$\approx$60\,000 in
the spectral range 3\,200 to 4\,600 {\AA}. After summing the spectra of
the same star, we achieved S/N  ratios per pixel ranging from about 50
to about 150. Due to the  higher apparent brightness of its stars, the
quality of Praesepe spectra is  remarkably high. The Hyades spectra we
used have  a resolution  of R$\approx$60\,000 and  a signal--to--noise
ratio per pixel ranging from 100 to 200, and from 20 to 30 at the core
of the Ca {\sc II} K line.

\section{Data analysis}
The  procedure  we  adopted   to  measure  chromospheric  activity  is
described in detail elsewhere \citep{paper1,letter}.  Here I present a
complete summary.

As an indicator of chromospheric activity we used $F^{\prime}_K$, i.e.
the energy flux of the Ca  {\sc II} K~line emitted per unit surface in
the chromosphere.  We also computed the value normalised to bolometric
emission:

$\log   R^{\prime}_K  =  \log   \left(   {F^{\prime}_K} / {\sigma
    T_{eff}^4} \right)$.

The spectra were normalised by  dividing their intensity by the counts
at the pseudo continuum point 3950.5--{\AA}.

The normalised flux was integrated over a 1--{\AA} wide region centred
on  the  Ca  {\sc  II}   K  line.   This  region  coincides  with  the
chromospheric emission  peak.  The result  of this integration  is the
1--{\AA}~K~index  and includes  a contribution  from  the photosphere.
The stars in NGC~3680, IC~4651,  M~67, IC~4756 and NGC~5822 showed the
interstellar  K  absorption  line  which  affected  the  chromospheric
K--line  feature.   As  for  NGC~3680,  IC~4651, and  M~67,  we  could
evaluate the contribution of the interstellar absorption, by observing
hot stars in each of these clusters, and correct for it.  This was not
possible, instead, for stars in  IC~4756 and NGC~5822.  To measure the
1--{\AA}~K~index  in these  stars, we  integrated the  normalised flux
only in the uncontaminated part  of their profile.  The Praesepe stars
are  unaffected  by  interstellar   absorption  thus  allowing  us  to
calculate a ratio between the flux measured over the 1--{\AA} and that
measured  over the  portion  of  the feature  which  is unaffected  by
interstellar line in all stars.  The initial measures for the affected
stars were  then multiplied by this  factor to give  a final corrected
1--{\AA}~K~index.   The  errors involved  in  the  measurement of  the
1--{\AA}~K~index were evaluated to be within 6\%.

For  the  Sun,  we  used  the 1--{\AA}~K~index  measurements  made  by
\cite{wl81}.   They monitored  solar chromospheric  activity  from the
first minimum to the maximum of the 21$^{st}$ solar--activity cycle.

Subtraction of  the photospheric contribution  to the 1--{\AA}~K~index
was  made as  follows.   For  the Sun,  we  computed the  photospheric
correction using a  solar--photosphere synthetic spectrum (courtesy of
P. Bonifacio).  For the other stars, the photospheric contribution was
computed by  scaling the solar  photospheric contribution by  a factor
that depends  on the stellar  parameters.  These scaling  factors were
computed using  the spectral  synthesis code of  MOOG \citep[][version
2002]{MOOG},  and  Kurucz's  grid  of models  \citep{kur93}.   Stellar
parameters,   namely   temperature,   gravity,   microturbulence   and
metallicity,    were   known   from    our   chemical    analyses   in
\cite{papoc1,papoc2} and, for Hyades stars, from \cite{paulson}.

In  order to transform  the 1--{\AA}~K~index,  which is  an equivalent
width, into  an intrinsic  flux, namely $F^{\prime}_K$,  we multiplied
the former by the flux at the stellar surface of the pseudo--continuum
radiation  at  3950.5--{\AA},  which  was obtained  from  the  stellar
temperatures using  relationships published in  \cite{sergio, luca85};
and \cite{luca88}.

The employment of stellar  temperatures from spectroscopic analysis of
iron lines  instead of published  colours avoids the use  of uncertain
reddening estimations.

\begin{figure*}
\includegraphics{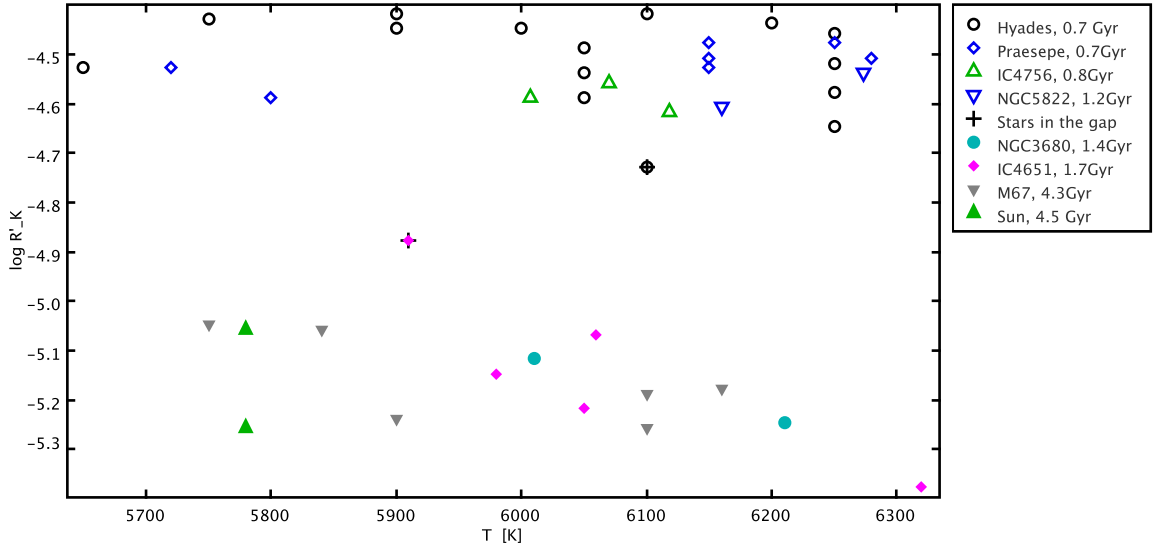}\\
\caption{%
  Diagram  of  CA  as  a  function  of  the  temperature  using  $\log
  R^{\prime}_K$  as a  proxy  for CA  .   For the  Sun, 2  datapoints,
  corresponding to the  maximum and the minimum of  an activity cycle,
  are  plotted.  Datapoints relative  to stars  above Vaughan--Preston
  gap  are plotted  with  empty symbols,  datapoints  below with  full
  symbols. The typical $1 \sigma$ error is 0.1 dex on the ordinate and
  110 K on  the abscissa.  The ages indicated in  the legend are taken
  by \cite{sal}.  } 
\label{fig}
\end{figure*}

The results  of the  reviewed analysis  of the old  sample and  of the
analysis of the  new sample, published in \cite{letter},  were used to
produce the diagram of CA as a function of temperature shown in Figure
\ref{fig}.

\section{Discussion.}
It was already pointed out  back in the eighties that the distribution
of CA  is markedly  bimodal \citep{vpgap}. Namely  very few  stars are
less active  than the  Hyades and significantly  more active  than the
Sun. This  underpopulated range  of values is  usually referred  to as
Vaughan--Preston (VP) gap. \cite{hartmann}  explained it as a combined
effect of  CA saturation in active stars  and a basal level  in the CA
indicator  used by VP,  due to  the photospheric  flux. These,  it was
claimed, enhance  the impression of a  gap.  A variation  in the local
stellar birthrate was also invoked.  Other studies consider the VP gap
a  result of  the  nature of  the  CA evolution  \citep[e.g.][]{dmr81,
  middelk}.  Our Figure  \ref{fig} corroborates the latter hypothesis:
all stars  younger than 1.2  Gyr, except one,  lie above the  gap, all
stars older than 1.4 Gyr with  the exception of one, lie below it,  
indicating a drop of CA level in a very short time. The two exceptions 
are probably due  to a particular phase of the activity cycle.   It is 
worth noticing  that the stars in either side of
the  gap differ  not only  in CA  level, but  also in  its trend  as a
function  of temperature: $\log  R^{\prime}_K$ (or  equivalently $\log
R^{\prime}_{HK}$) depends  weakly on  the temperature for  stars above
the gap,  while it  has a distinct  decreasing trend for  stars below.
This can  be seen  from our Figure  \ref{fig}, and it  appears clearer
when a larger temperature range  is considered, like, for instance, in
\citeauthor{MH} (\citeyear{MH},  see Figure 4  therein).  Furthermore,
short--term  temporal variations  of CA  are large  and  irregular for
active stars  and small and regular for  inactive ones \citep{cycles}.
There  must be  a major  event at  a given  time of  the  stellar main
sequence life time, that  changes the way radiative heating mechanisms
occur in the chromosphere, and,  as a consequence, their dependence on
stellar parameters as well as the shape and length of CA cycles.

\cite{fawzy}  present  theoretical  calculations  that  reproduce  the
observed  trend  of CA  with  stellar  temperature.   They invoke  two
heating mechanisms: the magnetic--wave and the acoustic--wave heating.
Theoretical chromospheric  fluxes for  the former mechanism  match the
observed flux  of the active  stars, while theoretical fluxes  for the
latter mechanism  match the observed  flux of the inactive  stars.  In
order to provide  a physical explanation for the  occurrence of the VP
gap, \cite{dmr81}  proposed a transition  from a complex to  a simpler
magnetic--field morphology which occurs  at the time when the rotation
decreases  enough   to  reach  a  threshold   value.   More  recently,
\cite{barnes}  detected two  sequences  in the  period--versus--colour
diagram of  open clusters, and  he associated them with  two different
rotation  morphologies,  intertwined  with  stellar  magnetic  fields.
\cite{bv07} suggested a change of dynamo mechanism to explain the fact
that  stars   occupy  two  very  distinct  sequences   in  a  rotation
period--versus--cycle period  diagram. Not all  of these works  can be
used  to  explain our  results.   What  is  relevant for  the  present
discussion is the possibility of a  change in the nature of the dynamo
mechanism  taking place  at  a specific  point  of the  main--sequence
lifetime of solar--type stars.

The  most important  conclusion of  our analysis  is that,  before and
after the fast drop  of CA from above to below the  VP gap in only 200
Myr, its alleged  smooth decay must be much  less important than short
term  variations.   We  noted   first  in  \cite{paper1}  that  CA  in
intermediate   age  clusters   had  already   dropped  to   the  solar
level. After  the reanalysis of the data  using temperature determined
spectroscopically from  iron lines, 1  out of the  7 stars in  the two
intermediate age  clusters IC~4651 and  NGC~3680 turned out to  lie in
the VP  gap.  In addition, the membership  of one of the  two stars in
NGC~3680 has  been questioned  \cite{chato}.  However, the  data still
point undoubtedly to a lack of evolution after 1.4 Gyr, since 5 out of
6 secure intermediate age stars have CA levels equal or lower to those
spanned by the Sun and by M~67 stars (empty dots in Figure \ref{fig}).
This conclusion is also corroborated  by the work of \cite{lyra}.  The
other  part  of the  conclusions,  that  regarding  the time  interval
between  0.7  and 1.2  Gyr  (filled  dots  in Figure  \ref{fig}),  was
achieved  after the  analysis  of  5 stars  in  NGC~5822 and  IC~4756,
together with  the older data  on Hyades and  Praesepe \citep{letter}.
We could show that metallicity is unlikely to play a major role, since
there is  no significant difference  between three almost  coeval open
clusters   at  different   metallicities:  Praesepe,   at  [Fe/H]=0.27
\citep{papoc1}, Hyades, at [Fe/H]=0.13 \citep{paulson}, and IC~4756 at
[Fe/H]=0.01 \citep{papoc2}.  The quoted iron abundance for Praesepe is
somewhat    different   from    other   published    works   \cite[see
e.g.][]{chato}.  However,  this result  is based on  the high--quality
spectra described in Section  \ref{datasample}, the iron abundance was
obtained  adopting both  spectroscopic  and photometric  temperatures,
several ways  of normalizing  for the continuum  were tried,  and even
different people performed EW  measurements in order to avoid personal
bias, and in any case  the result was [Fe/H] significantly higher than
0.2 dex.

Our  conclusions on  CA evolution  are still  based on  few  stars per
cluster, and  they partly rely on  the age scale  of \cite{sal}, which
uses  a  calibration  of  a  photospheric age  indicator  made  on  11
clusters.  In order to strengthen our results, or to disprove them, we
need more CA measurements in  solar--type stars in open clusters based
on high--resolution spectra and  direct age determinations of the open
clusters used based on  improved colour--magnitude diagrams. We should
both reobserve the already  analysed stars, to obtain a time--averaged
CA level, and choose other targets.

The scenario suggested  here contradicts the common belief  that CA is
well correlated  with age, however our investigation  does not include
stars younger  than the Hyades, some  of which are  indeed more active
than the  Hyades stars.   In addition, the  fact that young  stars are
active and old  stars are not is not questioned here,  and it causes a
weak correlation between CA and age.  We believe that data used in the
literature to prove a deterministic relation between CA and age do not
disprove our  alternative view,  and that more  data are  necessary to
achieve a definitive conclusion.

For  instance  \cite{sod91}  used  Mount--Wilson  data  about  several
solar--type stars  in visual binaries  and single F dwarfs,  with ages
from Str\"{o}mgren photometry in  order to prove that the relationship
between  CA and  age  is,  using their  words,  deterministic and  not
statistical.  Figure  3 in the  aforementioned paper shows  that $\log
R^{\prime}_{HK}$  and  the logarithm  of  age  are indeed  correlated.
However we notice, analysing data in Table 1 and 2 therein, that among
stars with $\log  R^{\prime}_{HK} > -4.9$, i.e.  more  active than the
Sun,  the correlation between  $\log R^{\prime}_{HK}$  and age  is not
significant, the Pearson correlation coefficient is about 0.1.  As for
the other stars,  i.e.  those with $\log R^{\prime}_{HK}  < -4.9$, the
Pearson coefficient is 0.35, and  its sign is positive, unlike what is
expected according  to an  activity decay.  In  either group  of stars
there   is,   instead,  a   significant   correlation  between   $\log
R^{\prime}_{HK}$ and the B-V colour, and it has opposite signs.

Another   calibration  of   CA  evolution   with  time   is   that  of
\cite{lach99}. From Fig.  4 therein,  it can be seen that their result
agrees with ours as far as inactive stars are concerned, i.e.  CA does
not evolve  after it has crossed the  VP gap.  As far  as active stars
older than  the Hyades  are concerned, there  are 4 to  6 data--points
(for  two stars it  is not  possible to  say whether  or not  they are
younger than the  Hyades and therefore out of  the range considered in
the present study).  For this group, the Pearson coefficient indicates
a level  of anticorrelation between age and  $\log R^\prime_{HK}$ that
is  weak (-0.27)  or  fair (-0.57),  depending  on how  many stars  we
consider.

From \cite{MH} it  is clear that the one open  cluster that supports a
strict  monotonicity  of CA  time  evolution  (once short--time  scale
variations  are smoothed out)  in the  age range  between that  of the
Hyades  and  that of  M~67  and the  Sun,  is  NGC~752. This  cluster,
according  to \cite{sal},  is  older than  NGC~5822  and younger  than
NGC~3680, i.e.  exactly in the range in which we expect the transition
to  occur. However,  \citeauthor{MH} also  show that  companions  of a
binary system tend  to have similar CA level,  especially if they have
similar colours.   According to the assumption  that CA is  a good age
indicator, this circumstance is easily  interpreted as due to the fact
that companions  of a  binary system are  coeval.  We suggest  that it
could be instead related to the nature of the binary systems.

We conclude  that data  in \cite{sod91,lach99,MH} are  compatible with
the conclusion that, within the age range from 0.7 to 1.2 Gyr and from
1.4 Gyr  to solar age,  any age--activity relationship must  be weaker
than the short term variations

\acknowledgments  I thank the  reviewers for  useful comments  and the
organization of the HELAS--NA5  workshop for having offered lodging in
the venue of the conference.  I acknowledge the support of the Funda\c
c\~{a}o para a Ci\^{e}ncia e a Tecnologia (Portugal) through the grant
SFRH/BPD/39254/2007 and the project PTDC/CTE-AST/65971/2006.


\end{document}